\documentstyle[12pt]{article}
\newcommand{\bce}{\begin{center}} 
\newcommand{\ece}{\end{center}}
\newcommand{\beq}{\begin{equation}}
\newcommand{\eeq}{\end{equation}}
\newcommand{\bea}{\vspace{0.25cm}\begin{eqnarray}}
\newcommand{\eea}{\end{eqnarray}}

\newcommand{\ba}{\begin{array}}
\newcommand{\ea}{\end{array}}

\newcommand{\doublespace}{
    \renewcommand{\baselinestretch}{1.6}\large\normalsize}

\def\lsim{\mathrel{\rlap{\lower4pt\hbox{\hskip1pt$\sim$}}
    \raise1pt\hbox{$<$}}}         
\def\gsim{\mathrel{\rlap{\lower4pt\hbox{\hskip1pt$\sim$}}
    \raise1pt\hbox{$>$}}}         

\def\beq{\begin{equation}}
\def\endeq{\end{equation}}
\def\bea{\begin{eqnarray}}
\def\arr{\begin{eqnarray}}
\def\eea{\end{eqnarray}}

\def\q2{$Q^{2}$}
\def\s2{2$S$}
\begin{document}
\doublespace 

\vspace*{-2cm}
\begin{flushright}
{\large\bf
KFA-IKP(TH)-1997-17\\
30 July 1997~~~}
\end{flushright}
 
\bigskip
 
\begin{center}
  
  {\Large\bf
Absorptive corrections to the one pion exchange and 
measurability of the small-$x$ pion structure function
at HERA
}
 
\vspace{1.5 cm}
 
  {\Large N.N.Nikolaev$^{a,b}$, J.Speth$^{a}$ and  B.G.Zakharov$^{b}$ }
  \bigskip
  \bigskip

{\it $^{a)}$Institut  f\"ur Kernphysik,
        Forschungszentrum J\"ulich,\\
        D-52425 J\"ulich, Germany\medskip\\
 $^{b)}$L. D. Landau Institute for Theoretical Physics,
        GSP-1, 117940,\\ ul. Kosygina 2, 117334 Moscow, Russia
        \medskip\\}

\vspace{.5cm}

  {\bf\large Abstract}
\end{center}
We compare the absorptive corrections to the one pion exchange
in $pp\rightarrow Xn$ and $ep\rightarrow e'Xn$ reactions.
It is shown that the absorption is stronger
in the case of $pp$ collision. The difference in strength of 
the absorption for 
the $pp\rightarrow Xn$ and $ep\rightarrow e'Xn$
reactions breaks the factorization peculiar 
to the pure one pion exchange.  
We evaluate the emerging model dependence of extraction of the
small-$x$ pion structure function from an analysis of the HERA data on
the neutron production at physical values of $t$.

\vspace*{3cm}
 
\noindent

 

\newpage



\section{Introduction}

The idea of using pions from the pion cloud of the proton as targets 
for measuring cross sections of interaction of different projectiles 
with pions goes back to Chew and Low \cite{ChewLow}. For instance, 
the inclusive reaction $ap\rightarrow nX$ can be viewed as a breakup 
of the $\pi^{-}n$ Fock state of the physical nucleon when the projectile 
$a$ interacts with the $\pi^{-}$. It has been well established 
\cite{ISR,Hanlon1,Eisenberg,Hanlon2} 
that the pion exchange of Fig.~1a is the dominant mechanism of this 
inclusive reaction in the region of small transverse momenta 
$p_{\perp}^{2}\lsim 0.2-0.3$ GeV$^{2}$ and $z\sim $0.7-0.9, where 
$z=p_{z}^{c.m.}/p_{max}$ is the neutron Feynman variable. The early 
discussions \cite{Bishari,Pumplin,Kaidalov1} of this reaction focused 
on the so-called triple-Regge 
formalism which is appropriate at high energies and large 
values of the Regge parameter, 
${1\over {1-z}} >> 1$. In terms of the inclusive structure function,
$$ 
f(z,p^{2})=\frac{z}{\pi}\frac{d\sigma}{dz dp^{2}}\,\,,
$$
the pion exchange contribution in the triple-Regge approach reads
\beq
f_{\pi}^{a}(z,\vec{p})=\frac{g_{pn\pi}^{2}}{2(2\pi)^{3}}
\frac{|t|}{(t-m_{\pi}^{2})^{2}}
F^{2}(t)(1-z)^{1-2\alpha_{\pi}^{'}(t-m_{\pi}^{2})}
\sigma_{tot}^{a\pi}(s_{X})
\,\,.
\label{eq:1}
\eeq
Here $g_{pn\pi}^{2}/4\pi=27.5$ \cite{gpnpi1}, $\sigma_{tot}^{a\pi}$ is
the $a\pi$ total cross section, $s$ and  $s_{X}=s(1-z)$ is the $ap$ and
$a\pi $ 
center of mass energy squared, $t$ is the pion momentum squared,
$\alpha_{\pi}^{'}$ is the slope of the pion Regge trajectory, 
$\alpha_{\pi}(t)=\alpha_{\pi}'(t-m_{\pi}^{2})$, and $F(t)$ is the 
form factor taking into account the off-shell effects. If $s_{X}>> 1$
GeV$^{2}$, { i.e.,} if $\sigma_{tot}^{a\pi}(s_{X})$ can be described
by the pomeron exchange, then $f_{\pi}^{a}(z,p^{2})$ is described by the
triple-Regge diagram $\pi\pi P$ shown in Fig.~1b.

The salient feature of Eq.~(\ref{eq:1}) is the factorization relation 
\beq
\frac{f_{\pi}^{a_{1}}(z,p)}{f_{\pi}^{a_{2}}(z,p)}=
\frac{\sigma_{tot}^{a_{1}\pi}}{\sigma_{tot}^{a_{2}\pi}}\,\,\,,
\label{eq:2}
\eeq
which for $a_{1}=\pi,K$ and $a_{2}=p,n$ has been used in practice 
for determination of the $\pi\pi$ and $K\pi$ total cross sections 
\cite{Hanlon1}. As a
matter of fact, the factorization relation (\ref{eq:2}) holds 
for both the reggeized pion exchange at $1-z \ll 1$ and the elementary
pion exchange at somewhat smaller $z$. 
On the other hand, because the $\pi N$ total cross sections are
known from direct measurements, one can use the $pp\rightarrow Xn$ 
reaction to fix the magnitude and the $t$-dependence of the form 
factor $F(t)$. Then, this form factor $F(t)$ can be used to
extract $\sigma_{tot}^{\pi(K)\pi}$ from the experimental data on 
the $\pi(K)p\rightarrow Xn$ reactions using (\ref{eq:1}). The crucial
point about the factorization relation is that at fixed $s_{X}$ in 
the  r.h.s., the l.h.s. of Eq.~(\ref{eq:2}) must not depend on $z$.

By extension of the factorization relation (\ref{eq:2}) to the real
and/or virtual photons, $a=\gamma,\gamma^{*}$, one may hope to determine
the cross section of real and virtual photoabsorption on pions. In
the latter case, there emerges a possibility of measuring the pion
structure function \cite{KFA},
$F_{2}^{\pi}(x_{\pi},Q^{2})=Q^{2}\sigma_{tot}^{\gamma^{*}\pi}(x_{\pi},Q^{2})
/4\pi^{2}\alpha_{em}$,
at very low $x_{\pi}={x\over 1-z}$, unaccessible in the Drell-Yan experiments. 
The measurements of the semi-inclusive $ep\rightarrow e'Xn$ cross section 
are now in progress at HERA \cite{HERA1}, and evaluation of the accuracy
of the factorization relation (\ref{eq:2}) is called upon.

On of limitations on 
the accuracy of the factorization relation is backgrounds to 
the pion exchange.
The background contributions connected with production
of $\pi n$ states through the one pion exchange (Fig.~2) 
and heavy meson exchanges were estimated in 
the recent works \cite{KFA,BZK}. 
The results of Refs. \cite{KFA,BZK} show that in the region
$z \sim 0.8$ and $p^{2}_{\perp}\lsim 0.2-0.3$ GeV$^{2}$ these mechanisms
give a relatively small ($\sim 10-20$\%) background.
This estimate is in a qualitative agreement with the earlier
analysis \cite{Kaidalov1},
which 
gives $\sim 20$\% background for $z\sim 0.8$ at $|t|\sim 0.1-0.2$ GeV$^{2}$.
%
%
%
At high $s_{X}$ both the $\pi\pi P$ and background contributions 
are proportional to the $\gamma^{*}\gamma^{*} P$ coupling. For this reason,
at HERA energies, even the 10-20\% background will give 
only a negligible violation of the factorization relation (\ref{eq:2}).
Therefore, one could have concluded that the uncertainties 
of the determination of the pion structure function from the semi-inclusive 
$ep\rightarrow e'Xn$ data will not exceed  a few percent.
This would have made
the $ep\rightarrow e'Xn$ reaction competitive with the $\pi N$ Drell-Yan 
process, the interpretation of which presently involves  theoretical 
uncertainties $\sim 10$ per cent  because of the so-called K-factor 
\cite{Drell-Yan}.

Unfortunately, the real situation is more complicated due to the absorption 
corrections to the pion exchange mechanism
generated by the double reggeon pion-Pomeron exchange shown in Fig.~3.
peculiar to 
the pure pion exchange mechanism.
%
The diagram of Fig.~3 takes into account the initial and final state
interaction effects. The absorption is known to suppress considerably 
the pion pole contribution in hadronic inclusive reactions. Typical
estimates for the absorption factor are $K_{abs}=(f_{\pi}+f_{abs})/f_{\pi}
\sim 0.4-0.7$, where $f_{abs}$ is the absorptive correction 
\cite{ABS1,ABS2,ABS3,ABS4}. The important finding is that the absorption 
corrections are approximately the same for $pp$ and $\pi(K)p$ collisions. 
For the reference reaction $p\rightarrow n$ reaction 
the effect of the absorptive $K_{abs}$-factor can be approximately 
included into the absorption-modified  off-shell form factor, $F_{abs}(t)$.
Then, if one takes this absorption-modified form factor from the
$pp$ data, there will be only marginal corrections
to the $\pi(K) p$ cross section determinations based on 
the triple-Regge formula (\ref{eq:1}) \cite{ABS5}. However, as we shall
argue in the present paper, one must expect substantial reduction of 
the absorption strength from hadrons to virtual photons. For this
reason, one cannot use the effective form factor $F(t)$ adjusted to the
description of $ap\rightarrow nX$ reactions for an analysis of the
$ep\rightarrow e'nX$, because the so obtained values of $\sigma_{tot}
^{\gamma^{*}p}$ will be overestimated by a factor 
$R(\gamma^{*}/p)=K_{abs}(\gamma^{*}p\rightarrow Xn)/K_{abs}(pp\rightarrow Xn)$.
In the present paper we perform comparative analysis of absorption 
in $ap\rightarrow nX$ and $ep\rightarrow e'nX$ reactions, and
estimate the model dependence of the determinations of $\sigma_{tot}
^{\gamma^{*}\pi}$ because of the theoretical uncertainties in the 
absorptive corrections.

%

Evidently, absorption of the projectile hadron $a$  and final state 
$X$ is strong for impact parameters $b \lsim R_{p}$, where $R_{p}$ is 
the proton radius. The size of the pion cloud around 
nucleon, $\lsim 1/m_{\pi}$, is comparable to the radius of the proton 
$R_{p}$. Consequently, the pure pion exchange must be considerably 
modified by the absorption.
%
%
Unfortunately, at present,  a rigorous treatment of the absorptive effects
generated by the diagram of the type in Fig.~3 is impossible.  
In the literature the absorptive effects in the triple-Regge
region is commonly described in the framework of Reggeon
calculus. The corresponding reggeon diagrams for the pion
exchange mechanism are shown in Fig.~4. 
This approach is motivated by the generalization of the AGK cutting 
rules \cite{AGK}, derived within $\lambda \phi^{3}$ field theory,
to inclusive reaction in the triple-Regge regime \cite{Capella}.
A nontrivial consequence of the AGK rules is that, after summing over 
the final states,  all the initial and final states interaction effects 
in the inclusive cross section can be described by the triple-Regge 
diagrams with additional Pomeron exchanges depicted in Fig.~4, and by 
the corresponding enhanced diagrams containing the triple-Pomeron 
coupling $r_{PPP}$.
The latter are usually neglected due to smallness of the $r_{PPP}$.
The major absorptive effect comes from the graphs of Figs.~4a,b, which
correspond to interference of the $\pi$ (Fig.~1a)  and $\pi P$ (Fig.~2)
exchanges.
The diagram of Fig.~4c related to the $\pi P$ exchange amplitude
squared gives a relatively small positive contribution to the
inclusive cross section.  

For hadronic $ap\rightarrow X n$ reaction the contribution of 
the graphs shown in Fig.~4  
in the quasieikonal  approximation 
is given by
\bea
f_{abs}^{a}(z,\vec{p}_{\perp})=
\frac{i C_{1}}{8\pi^{2}s}\int d\vec{k} \,T_{ap}(\vec{k})
f_{\pi}^{a}(z,\vec{p}_{\perp},\vec{k},0)-
\frac{i C_{1}}{8\pi^{2}s}\int d\vec{k}\,
f_{\pi}^{a}(z,\vec{p}_{\perp},0,\vec{k})\,T_{ap}^{*}(\vec{k})\nonumber\\
+
\frac{C_{2}}{(8\pi^{2}s)^{2}}\int d\vec{k}_{1}
d\vec{k}_{2}\, T_{ap}(\vec{k}_{1})
f_{\pi}^{a}(z,\vec{p}_{\perp},\vec{k}_{1},\vec{k}_{2})\,
T_{ap}^{*}(\vec{k}_{2})\,\,,
\,\,\,\,\,\,
\label{eq:3}
\eea
where 
$T_{ap}$ stands for the amplitude of elastic $ap$ scattering
(we use normalization $\mbox{Im}T_{ap}(\vec{k}=0)=s\sigma_{tot}^{ap}$),
$C_{1,2}$ are the shower coefficient for the diagrams of Figs.~4a,b
and Fig.~4c, respectively. They are introduced to 
take into account the inelastic intermediate states in
the $a\rightarrow a$ and $p\rightarrow n$ reggeon vertices.
$C_{1,2}=1$ in the eikonal approximation, when only elastic intermediate 
states are included.
The generalized $\pi\pi P$ structure function 
$f_{\pi}^{a}(z,\vec{p}_{\perp},\vec{k}_{1},\vec{k}_{2})$ for nonzero initial 
proton transverse 
momenta $\vec{k}_{1,2}$ 
appearing in Eq.~(\ref{eq:3}) is given by
\bea
f_{\pi}^{a}(z,\vec{p}_{\perp},\vec{k}_{1},\vec{k}_{2})=
\frac{g_{pn\pi}^{2}}{2(2\pi)^{3}}
\left[|t_{{min}}|+\frac{(\vec{p}_{\perp}-z\vec{k}_{1})
(\vec{p}_{\perp}-z\vec{k}_{2})}{z}
\right]
\frac{F(t_{1})}{(t_{1}-m_{\pi}^{2})}\,
\frac{F(t_{2})}{(t_{2}-m_{\pi}^{2})}\nonumber\\
\times
(1-z)
\exp[\Lambda (t_{1}-m_{\pi}^{2})+
\Lambda^{*} (t_{2}-m_{\pi}^{2})-\Lambda_{a\pi }(\vec{k}_{1}-\vec{k}_{2})^{2}]
\,\sigma_{tot}^{a\pi}(s_{X})
\,\,,\,\,\,
\label{eq:4}
\eea
where
$$
\Lambda=\alpha_{\pi}^{'}\left[\log\frac{1}{(1-z)}-\frac{i\pi}{2}\right]\,,
$$
$$
t_{i}=t_{{min}}-\frac{(\vec{p}_{\perp}-z\vec{k}_{i})^{2}}{z}\,,
$$
$$
t_{min}=-m_{p}^{2}\frac{(1-z)^{2}}{z}\,.
$$
In arriving at the formula (\ref{eq:4}) we have used the Gaussian
parameterization of the amplitude of elastic $a\pi $ scattering,
$\mbox{Im}T_{a\pi }(k)=
s_{X}\sigma_{tot}^{a\pi }\exp(-\Lambda_{a\pi }k^{2})$.


The numerical parameters for the $pp\rightarrow Xn$ reaction were fixed 
as follows. The shower coefficients $C_{1,2}$ can be written in the 
factorized form $C_{1}=C_{PP}C_{P\pi}$ and $C_{2}=C_{PPP}C_{P\pi}^{2}$. 
Here $C_{PP}(C_{PPP})$ and $C_{P\pi}$ are the shower coefficients
for the $a\rightarrow a$ double-Pomeron (triple-Pomeron) and 
$p\rightarrow n$ pion-Pomeron blobs in Figs.~4, respectively.
The shower coefficient $C_{PP}$ can be extracted from the data
on the diffractive process $ap\rightarrow a^{*}p$. For the proton
projectile, $a=p$, it has been found that $C_{PP}\approx 1.15$ 
\cite{Kaidalov2}. The $C_{PPP}$ was estimated in the two-channel approximation,
which yields $C_{PPP}\approx 1.09 C_{PP}$ \cite{ABS4}. The mass of the 
$\pi_{2}(1670)$ meson suggests $\alpha'_{\pi}=0.7$ GeV$^{-2}$. For
the elastic $pp$ scattering amplitude, which enters (\ref{eq:3}), we
take the standard Gaussian approximation, $T_{pp}(k)=(i+\rho)
s\sigma_{tot}\exp(-\Lambda_{pp}k^{2})$. The useful parameterization
for $\sigma_{tot}^{\pi^{-}p}(s_{X})$ is found in \cite{PPD}, for a good
compilation of the diffraction slope data see \cite{Kamran}. The real 
part of the $pp$ scattering amplitude is small, $\rho \ll 1$, and its
impact on absorption corrections is negligible. We take the 
Gaussian parameterization for the off-shell form factor,
$F(t)=\exp[R^{2}(t-m_{\pi}^{2})]$. The two free parameters, $R^{2}$ and 
$C_{P\pi}$, were fitted to the experimental data on the $pp\rightarrow Xn$ 
and $pn\rightarrow Xp$ reactions. We have used the ISR data \cite{ISR} 
on neutron production in $pp$ collision at $p_{\perp}=0$, and the results 
of Refs. \cite{Hanlon1,Eisenberg,Hanlon2} on the $p_{\perp}^{2}$-integrated 
cross sections for the $pn\rightarrow Xp$ reaction. We included in the 
fit the experimental points in the interval $0.7<z<0.9$. The contribution
of the background effects, which can give $\sim 10-30$\% of the experimental 
cross section at $z\sim 0.8$ \cite{Kaidalov1,KFA,BZK}, is modeled 
scaling the pion exchange contribution up by the factor 1.2. We are 
fully aware that this procedure oversimplifies the description of the 
background. However, because the experimental errors are substantial and
because the $z$ and $p_{\perp}$ dependence of the background is poorly
known theoretically, going after more sophisticated parameterizations of 
the background contribution is not warranted.

A fit to the above described data gives $R^{2}=-0.05\pm 0.08$ and 
$C_{P\pi}=0.67\pm 0.1$, with $\chi^{2}/N\approx 1.65$
Figs.~5, 6 show that the quality of the fit for in the region $z\sim 
0.7-0.9$ is good. As described above, the theoretical curves calculated 
with the fitted parameters $R^{2}$ and $C_{P\pi}$ are scaled up by the 
factor 1.2 to model the background contribution. The strength of 
absorption is seen in Fig.~7, in which we show by the $K_{abs}$-factor 
for $pp\rightarrow Xn$ reaction, calculated for $p_{\perp}^{2}=0,\,0.1,
\,0.2,\,0.3$ GeV$^{2}$  with the fitted parameters $R^{2}$ and $C_{P\pi}$.
Since the absorption only weakly depends on the energy, we show the 
results only for the incident proton momentum $p_{lab}=400$ GeV/c.
Fig.~7 shows that absorption is substantial and gets stronger for
smaller $z$, especially small $p_{\perp}^{2}$. This effect can be 
related to the decrease of the $\pi N$ spatial separation in the impact
parameter plane, $R_{\pi N}$, with the decreasing $z$. Indeed, the 
strength of the absorption is to a crude approximation proportional 
to the parameter $\sigma_{tot}^{pn}/(\Lambda_{pp}+R_{\pi N}^{2})$.
Because of non-vanishing $t_{min}$, the pion propagator takes the form 
\beq
{1 \over t-m_{\pi}^{2}}=
{-z \over p_{\perp}^{2} +z m_{\pi}^{2} + m_{p}^{2}(1-z)^{2}}\,.
\label{eq:Piprop}
\eeq
which gives an estimate 
\beq
R_{\pi N}^{2} \propto {1 \over z m_{\pi}^{2} + m_{p}^{2}(1-z)^{2}}\, .
\label{eq:RpiN2}
\eeq
The $K_{abs}$-factor decreases with the increase of the transverse 
momentum, which is naturally related to stronger absorption at small impact
parameters.

Strictly speaking, the applicability domain of the above triple-Regge 
is $\log {1\over 1-z} \gg 1$. The elementary pion exchange is more
appropriate for $\log {1\over 1-z} \lsim 1$. The considered region of 
$z=$0.7-0.9 is on the boundary between the reggeized and elementary 
pion exchanges and, as a matter of fact, the reggeization effects are
marginal. The transition  from the triple-Regge formulas to the
 light cone treatment of the elementary pion 
exchange as expounded in [10,24] is achieved the replacement 
$R^{2}\rightarrow R^{2}/(1-z)$ and by putting $\alpha'_{\pi}=0$. We
checked that such a light cone formalism with $R^2=0.19\pm 0.07$
GeV$^{-2}$, $C_{P\pi}=0.75\pm 0.1$ provides an equally viable
description of the $pp$ and $pn$ experimental data in the region 
$0.7<z<0.9$ ($\chi^2/N \approx 1.88$). Consequently, 
the specific Regge effects do not play any substantial role
in this kinematical domain.

Let us now consider the $\gamma^{*}p\rightarrow Xn$ reaction. We will 
estimate the  $K_{abs}$-factor for this case in two different ways. 
The first option is to extend to DIS the above outlined reggeon diagram
approach. As we shall see, in this case absorption is weak and the 
contribution from the diagram of Fig.~4c, which is quadratic in the
absorption amplitude of Fig.~3, can be neglected.
%
We evaluate the diagrams of Figs.~4a,b assuming that the ratio of the coupling 
of the Pomeron to the pion and proton equals 
$\sigma_{tot}^{\pi p}/\sigma_{tot}^{pp}\approx 2/3$. 
Then, the two-Pomeron blob in Figs.~4a,b 
can be expressed through  
the $\gamma^{*}p\rightarrow Xp$ diffractive cross section, 
$d\sigma_{D}^{\gamma^{*}p}/dk^{2}$, and the absorptive correction
to the pion exchange contribution can be written as 
\bea
f_{abs}^{\gamma^{*}}(z,\vec{p}_{\perp})=
-\frac{g_{pn\pi}^{2}}{3\pi^{3}}(1-z)C_{P\pi}
\mbox{Re}\frac{F(t)\exp[\Lambda^{*}(t-m_{\pi}^{2})]}{(t-m_{\pi}^{2})}
\int d\vec{k}\,\,
\frac{d\sigma_{D}^{\gamma^{*}p}}{d k^{2}}\,
\exp[-(\Lambda_{\pi p}-\Lambda_{pp})k^{2}]\nonumber\\
\times
\left[|t_{min}|+\frac{\vec{p}_{\perp}
(\vec{p}_{\perp}-z\vec{k})}{z}\right]
\frac{F(t^{'})\exp[\Lambda (t^{'}-m_{\pi}^{2})]}
{(t^{'}-m_{\pi}^{2})}\,\,,\,\,\,\,\,\,\,\,\,
\label{eq:5}
\eea 
where
$$
t'=t_{{min}}-\frac{(\vec{p}_{\perp}-z\vec{k}_{i})^{2}}{z}\,.
$$

For the $\gamma^{*}p$ diffraction cross we use the conventional 
Gaussian parameterization,
$d\sigma_{D}^{\gamma^{*}p}/dk^{2}=
\sigma_{D}^{\gamma^{*}p}B_{D} \exp(-Bk^{2})$,
here $\sigma_{D}^{\gamma^{*}p}$ is the total diffraction cross section.
We take $\sigma_{D}^{\gamma^{*}p}=\xi \sigma_{tot}^{\gamma^{*}p}$ with
$\xi=0.07$, which correspond to the results of the H1 \cite{H1} and 
ZEUS \cite{ZEUS1} experiments in the region $Q^{2}\sim 10-100$ GeV$^{2}$ 
and $x\sim 0.001$. For the diffraction slope we take 
$B_{D}=7$ GeV$^{-2}$ according to the measurements performed by  
the ZEUS Collaboration \cite{ZEUS2}.

The results for the $K_{abs}$-factor for this version of the absorption 
in the $\gamma^{*}p\rightarrow Xn$ reaction are shown in Fig.~7 by the 
dashed lines. The departure of the $K_{abs}$ from unity is much smaller
that for the $pp\rightarrow nX$ reaction, which is better seen in Fig.~8
where we show the ratio $R(\gamma^*/p)=K_{abs}(\gamma^{*}p\rightarrow Xn)
/K_{abs}(pp\rightarrow Xn)$. The departure of $R(\gamma^{*}p)$ from 
unity signals a strong factorization breaking.
Weak absorption for $\gamma^{*}p\rightarrow Xn$ 
is predicted because in the reggeon calculus the combined effect of the 
initial and final state interactions effects is described by the 
rescattering of the initial particles. In DIS the strength of initial
state rescattering is proportional to the small ratio $\xi =
\sigma_{D}^{\gamma^{*}p}/\sigma_{tot}^{\gamma^{*}p}\approx 0.07$. 
The counterpart of this parameter for the projectile proton is the ratio
$(\sigma_{el}^{pp}+\sigma_{D}^{pp})/\sigma_{tot}^{pp}\approx 0.25$, which
is by a factor $\sim 3-4$ larger because of the contribution of elastic
rescatterings. 

The caveat of the above reggeon calculus estimate is that the status 
of the AGK rules in QCD remains open. Its applicability is 
especially questionable in the $\gamma^{*}p\rightarrow Xn$ reaction. 
In QCD one can expect that the final state interaction effects in this 
reaction will be approximately as strong as in the $pp\rightarrow Xn$ 
reaction, and now we comment more on this option.
Indeed, the final state $X$, created in DIS after color exchange between 
the hadronic $q\bar{q}$ Fock component of the virtual photon and the pion,
looks like a color octet-octet system, $X_{DIS}=(q\bar{q})_{8}(q\bar{q})_{8}$.
In the $pp$ collision, similar color exchange between the proton and pion
creates $X_{pp}=(qqq)_{8}(q\bar{q})_{8}$. The both color octet-octet states
will have a similar transverse size, perhaps by a factor $\sim\sqrt{2}$
larger for the $X_{pp}$ state. Consequently, the strength of the final
state interaction of the state $X_{DIS}$ with the spectator neutron will
be as large as $\sim {1\over 2}$ of that of the state $X_{pp}$ in the
$pp$ collision.
 The initial state interaction in 
the $\gamma^{*}p$ case for the most part comes from the well known 
asymmetric $q\bar{q}$ configurations in the virtual photon light-cone wave 
function \cite{NZ}, which dominate the leading twist photon diffraction 
cross section. Precisely as in the above reggeon calculus considerations,
the absorptive effects for these $q\bar{q}$ configurations are suppressed 
by a factor $\sim 3-4$ as compared to those in the $pp$ collision. Thus, 
in QCD one can expect a significant enhancement of the absorptive effects 
in the $\gamma^{*}p\rightarrow Xn$ process as compared to prediction of 
the reggeon calculus.
For the numerical estimate of the absorptive $K$-factor for the 
$\gamma^{*}p\rightarrow Xn$ reaction in this, QCD motivated, version
one can use the formulas of the reggeon diagram approach for 
the $pp\rightarrow Xn$ reaction taking for the $C_{P\pi}$ a
value two times smaller than that for the $pp\rightarrow Xn$ reaction.
The results are presented in Figs.~7,~8 by the long-dashed lines and
show weaker factorization breaking. The difference between the 
$R(\gamma^{*}/p)$ for the two versions of absorption, 
and especially variations
of $R(\gamma^{*}/p)$ with $z$ and $p_{\perp}^{2}$, 
indicate the degree of model 
dependence of extraction of the pion structure function from the 
$\gamma^{*}p\rightarrow Xn$ data. Even in the  region $z\sim$ 0.7-0.9 
and $p_{\perp}^{2}\lsim 0.3$ GeV$^{2}$, which is optimal from the point 
of view of the dominance of the pion exchange, the uncertainties for 
the model dependence of absorption can be as large as $\sim 20$ per cent,
exceeding the potential background corrections.

In principle, all the problems with the parameterization of the off-shell
form factor and with the absorptive factor could have been eliminated
if an extrapolation to the Chew-Low unphysical point $t=m_{\pi}^{2}$
were possible.
It is interesting to find out whether the determination of
$\sigma_{tot}^{\gamma^{*}\pi}$ by such a Chew-Low extrapolation
is practically feasible. To this end 
we performed the following theoretical experiment. We approximate 
the $t$-dependence of the absorptive $K$-factor in
the region $p_{\perp}^{2}<0.1$ GeV$^{2}$ by a polynomial
$K_{abs}=a_{0}+a_{1}t+a_{2}t$. We checked that in the studied
range of $z$ an accuracy of such a parameterization is better than 1\%. 
Then, making use of 
this parameterization we extrapolate the absorptive $K_{abs}$-factors 
from the physical scattering domain, $t \leq t_{\min}$,
to the unphysical pion pole, $t=m_{\pi}^{2}$. 
In the ideal case,  the absorptive factor $K_{abs}=
(f_{\pi}+f_{abs})/f_{\pi}$ must extrapolate to unity at the pion pole.
The results are shown in Fig.~9. The smaller is the $z$, i.e., the 
larger is $|t|_{min}$, the poorer is extrapolation.
It is seen that in the case of the $\gamma^{*}p\rightarrow Xn$
reaction the extrapolated $K_{abs}$-factor equals 0.9-0.95 at $z\sim 0.8$,
while for the $pp\rightarrow Xn$ reaction it is $\sim 0.8$.
We conclude that the Chew-Low extrapolation is a very delicate
procedure, and it hardly can be used in practice for
a model independent 
extraction of $\sigma_{tot}^{\gamma^{*}p}$.

The detection of neutrons from interaction of the proton beam with 
the residual gas in the HERA ring, $pA\rightarrow nX$, provides a useful 
{\sl in situ} test of the performance of neutron detectors of ZEUS 
and H1 \cite{HERA1}. However, because of the intranuclear absorption 
and intranuclear rescatterings of beam proton and produced neutrons, 
the effective form factor $F_{A}(t)$ which one can deduce from the
$pA\rightarrow nX$, neither equals the form factor $F(t)$ for the
$pp$ reaction, nor can it be used as an input in the r analysis of
the $ep\rightarrow e'nX$ data. Consequently, the beam-gas interaction 
data cannot reduce the uncertainty in $R(\gamma^{*}/p)$.

To summarize, with the present state of the theory of  absorptive 
corrections, the above cited uncertainties in $R(\gamma^{*}/p)$ 
and, consequently, 
in the absolute normalization of the extracted pion structure function,
cannot be reduced. None the less, $\gamma^{*}p\rightarrow nX$ reaction 
will provide a useful information on the $x$ dependence of the pion 
structure function. For instance, if the $x$-dependence of the pion
structure function $F_{2\pi}(x,Q^{2})$ in the accessible region of
$x\lsim 10^{-4}$ is as strong as that of the proton structure function,
then this $x$-dependence cannot be masked by the uncertainties in the
evaluation of absorption corrections. On the large-$x_{\pi}$ end,
$x_{\pi}\sim$ 0.1., one can check a consistency with the Drell-Yan
data \cite{DY}. 


Finally, a comment on the impact of absorption effects on evaluations 
\cite{KFA,KFA2} of the mesonic corrections to the flavor content of
the proton structure function is in order. In these calculations, the
normalization of the form factor $F(t)$ has been deduced from the  
experimental data on $pp\rightarrow Xn$ reaction neglecting the 
absorption corrections. Because the absorption corrections are stronger
for the proton projectile than in DIS, such a simplified analysis
underestimates the effect of mesons in the proton structure function.
If the allowance for absorption is made, then the mesonic contributions
to the proton structure function will be enhanced by a factor 
$\sim R(\gamma^{*}/p)$. Here we wish to emphasize that because $t_{min}=-
{(m_{\Delta}^{2}-m_{p}^{2})(1-z)+m_{p}^{2}(1-z)^{2} \over z}$ for
the $pp\rightarrow \Delta X$ reaction is larger than for the $pp\rightarrow
nX$ reaction, the absorption corrections in the $\pi \Delta$ state are
stronger and the corresponding $R(\gamma^{*}/p)$ will be larger. This entails
the larger contribution from the $\pi \Delta$ Fock state to the proton 
structure function than estimated before. This effect can be of great 
importance from the point of view of the Gottfried sum rule violation
and the $\bar{u}$-$\bar{d}$ asymmetry in the proton, which are sensitive 
to delicate cancelation between the $\pi N$ and $\pi \Delta$ contributions 
\cite{KFA2}. An analysis of this problem with allowance for the absorption 
effects will be presented elsewhere. 

{\large \bf Acknowledgments}

 One of the authors (BGZ) would like to
thank IKP KFA for hospitality. This work was supported in part
by the INTAS grant 93-239ext.

\pagebreak

\pagebreak

{ \Large \bf Figure Captions}
\begin{itemize}
 
\item[Fig.1]
     The pion exchange amplitude for the $ap\rightarrow Xn$ 
reaction  (a) and the corresponding triple-Regge diagram for
the inclusive cross section (b).

\item[Fig.2]
     The pion exchange mechanism for the background 
$ap\rightarrow X\pi n$ reaction with production of the $n\pi$ system.

\item[Fig.3]
     The absorptive pion-Pomeron exchange amplitude for the 
$ap\rightarrow Xn$ reaction.

\item[Fig.4]
     The reggeon diagrams for absorption corrections to the $\pi\pi P$ 
diagram of Fig.~1b for the inclusive cross section.

\item[Fig.5]
     The $z$-distribution at $p_{\perp}=0$ for the $pp\rightarrow Xn$ 
reaction. The theoretical curves correspond to the contribution from
the absorption corrected pion exchange, scaled up by the factor 1.2
as described in the text. The experimental data from the ISR 
experiment \cite{ISR}.

\item[Fig.6]
     The inclusive cross section for the $pn\rightarrow Xp$ reaction
for different $|t|$ bins versus $z$ at: (a)  $p_{lab}=100$ GeV/c
\cite{Hanlon1} for $0.05<|t|<0.25$ GeV$^{2}$ (full circles) and 
$0.25<|t|<0.55$ GeV$^{2}$  (full quadrangles), (b) $p_{lab}=195$ GeV/c
\cite{Eisenberg} for $|t|<1.4$ GeV$^{2}$, 
 (c) $p_{lab}=100$ GeV/c
\cite{Hanlon2} for $|t|<1$ GeV$^{2}$, 
 (c) $p_{lab}=400$ GeV/c
\cite{Hanlon2} for $|t|<1$ GeV$^{2}$.
The theoretical curves are the same as in Fig.~6.

\item[Fig.7]
   The absorptive $K_{abs}$-factor versus $z$ for 
$p_{\perp}^{2}=0$ (a),
$p_{\perp}^{2}=0.1$ (b),
$p_{\perp}^{2}=0.2$ (c) and
$p_{\perp}^{2}=0.3$ (d) GeV$^{2}$.
 The solid lines correspond to the 
$pp\rightarrow Xn$ reaction at $p_{lab}=400$ GeV. The dashed lines
show the prediction of the reggeon calculus approach for
the $\gamma^{*}p\rightarrow Xn$ reaction at HERA energies, 
the long-dashed 
ones correspond to estimate of the $K_{abs}(\gamma^{*}p\rightarrow Xn)$
for the QCD motivated version of the absorption discussed in the text.

\item[Fig.8]
   The ratio 
$R=K_{abs}(\gamma^{*}p\rightarrow Xn)/K_{abs}(pp\rightarrow Xn)$
versus $z$.
The legend of boxes, i.e., values of $p_{\perp}^{2}$, and the legend of
curves are same as in Fig.~7.

\item[Fig.9]
   The polynomial extrapolation of the absorptive $K_{abs}$-factor 
to the Chew-Low point $t=m_{\pi}^{2}$ versus $z$. 
The legend of curves the same as in Fig.~7.

\end{itemize}
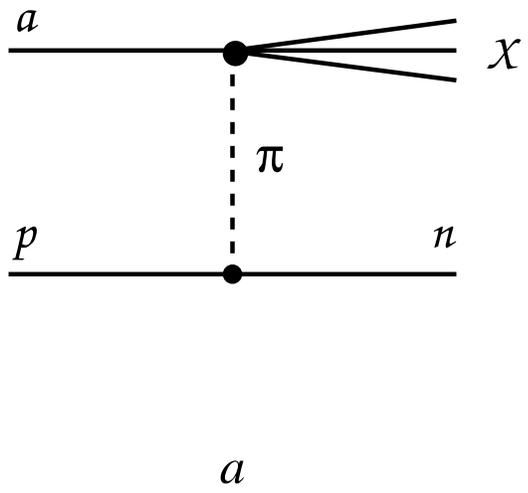
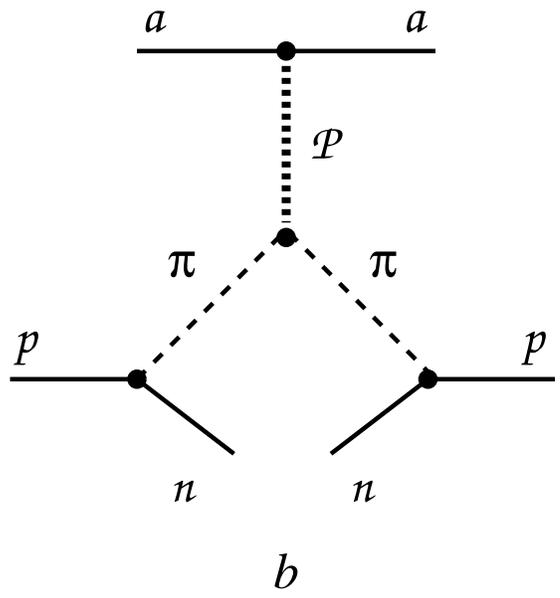
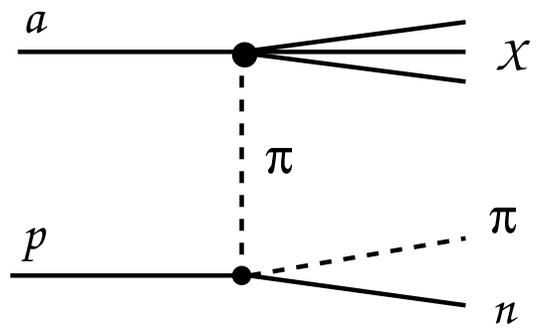
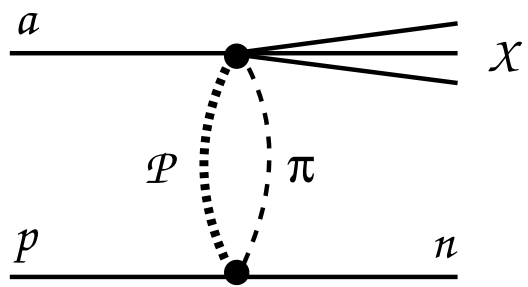
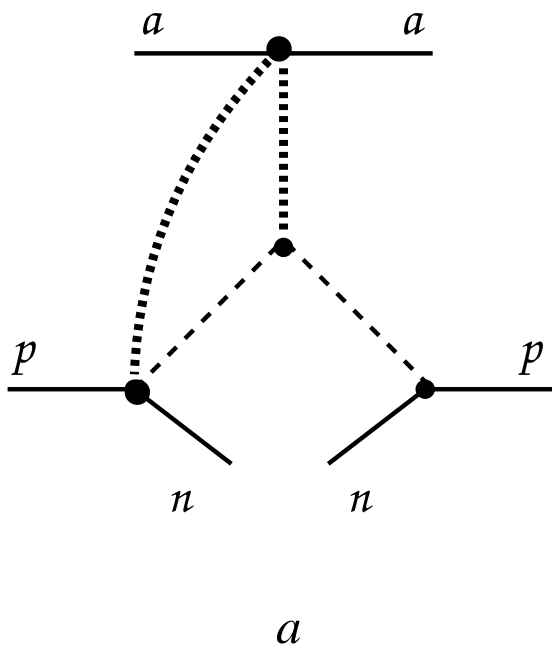
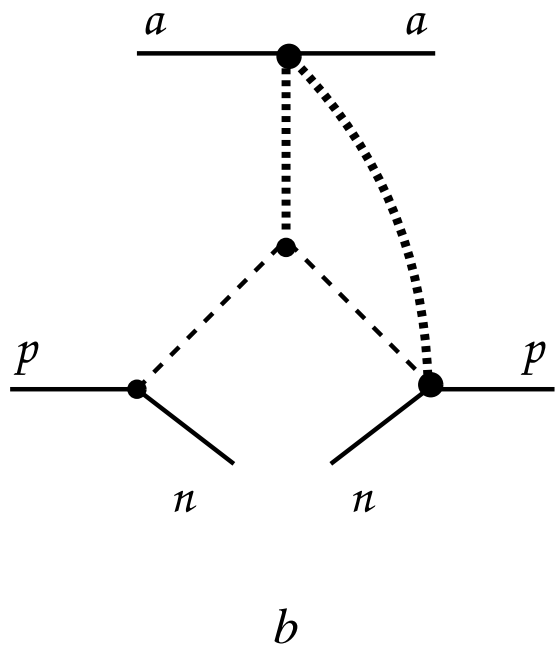
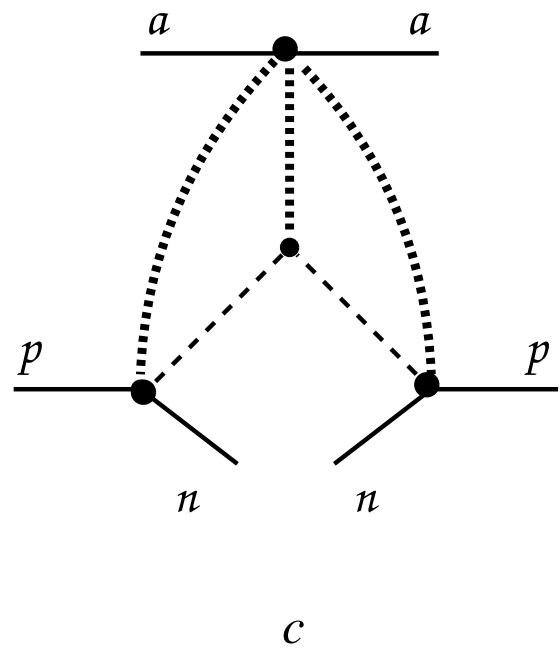
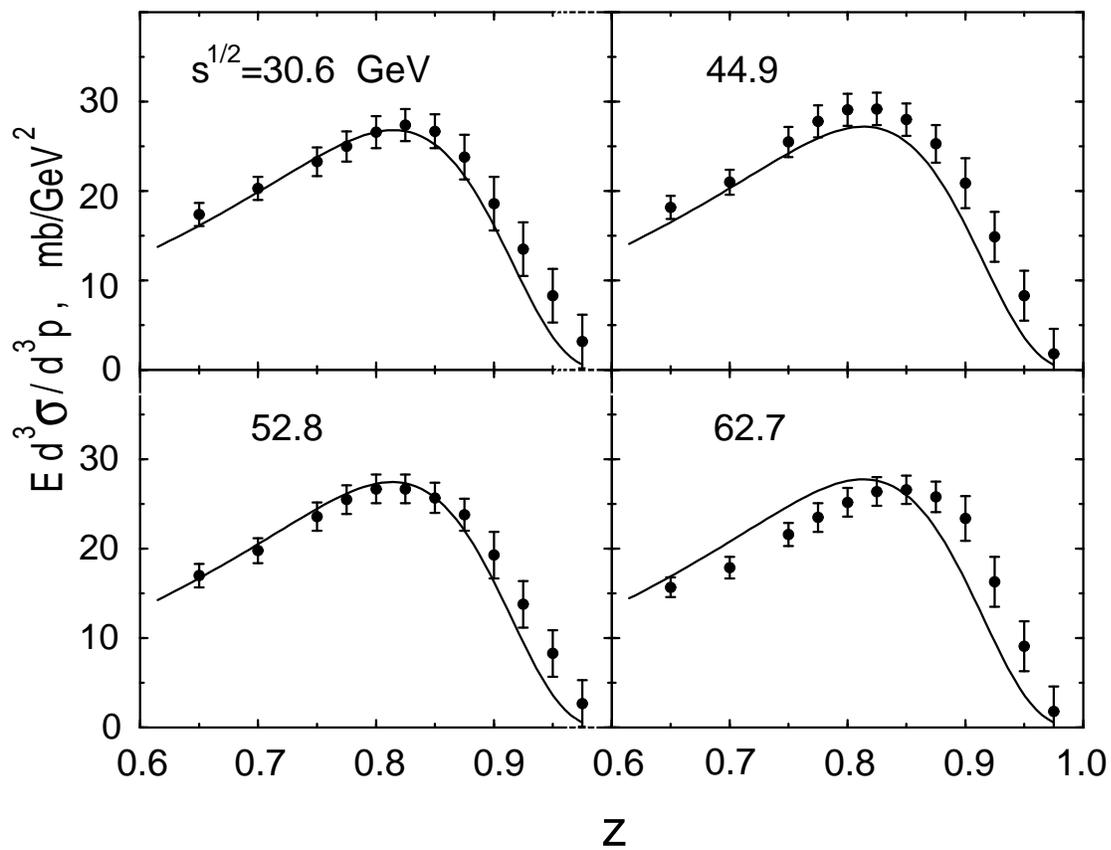
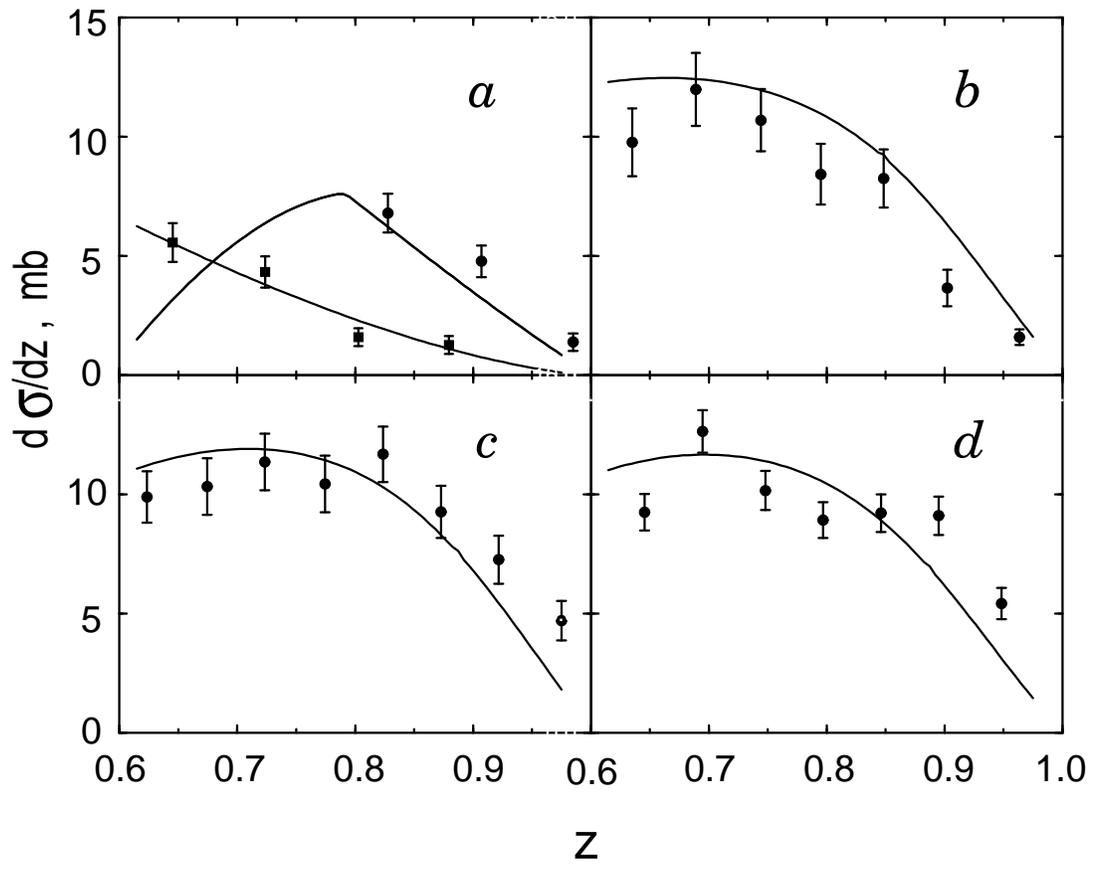
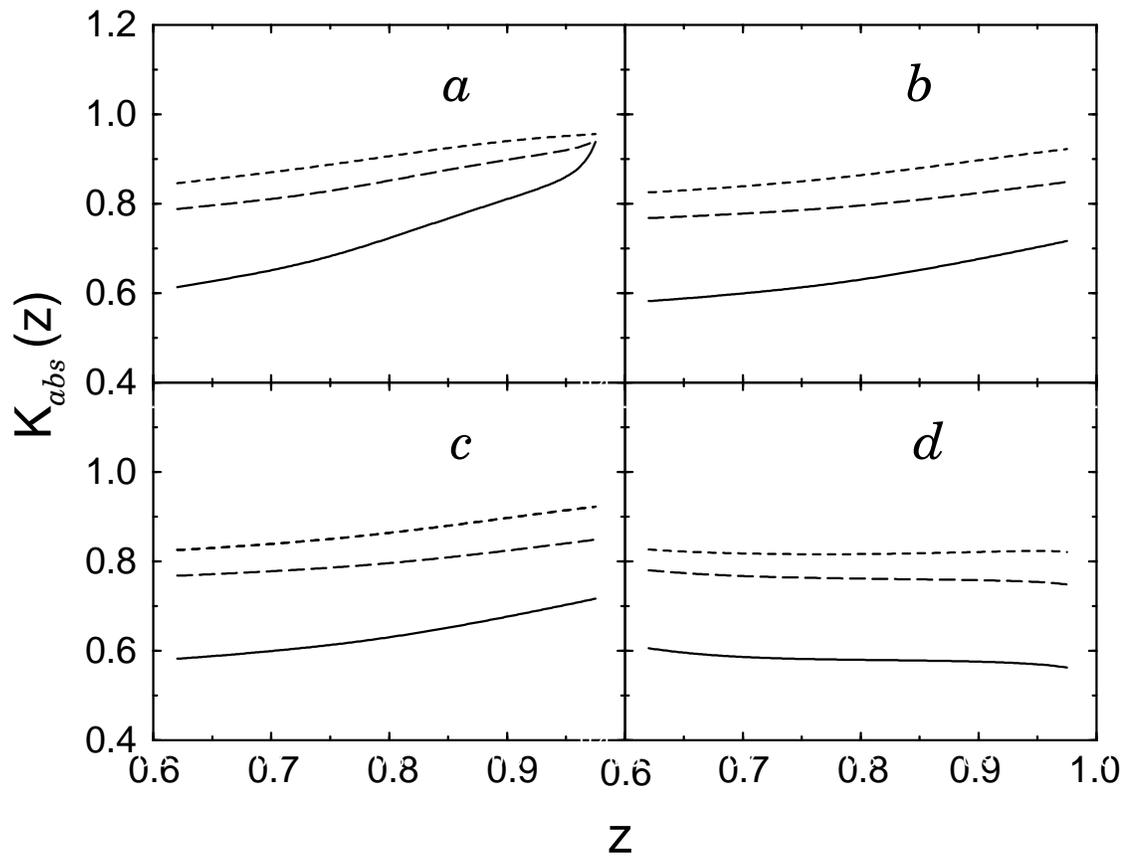
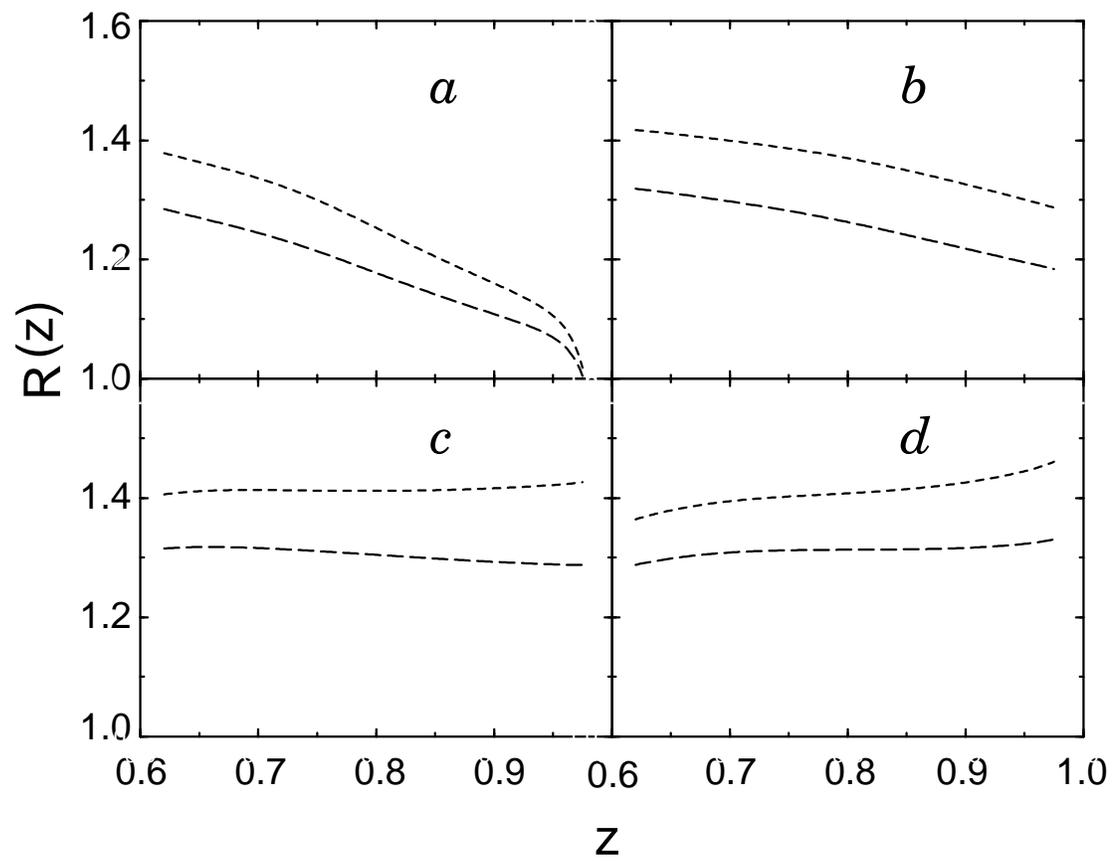
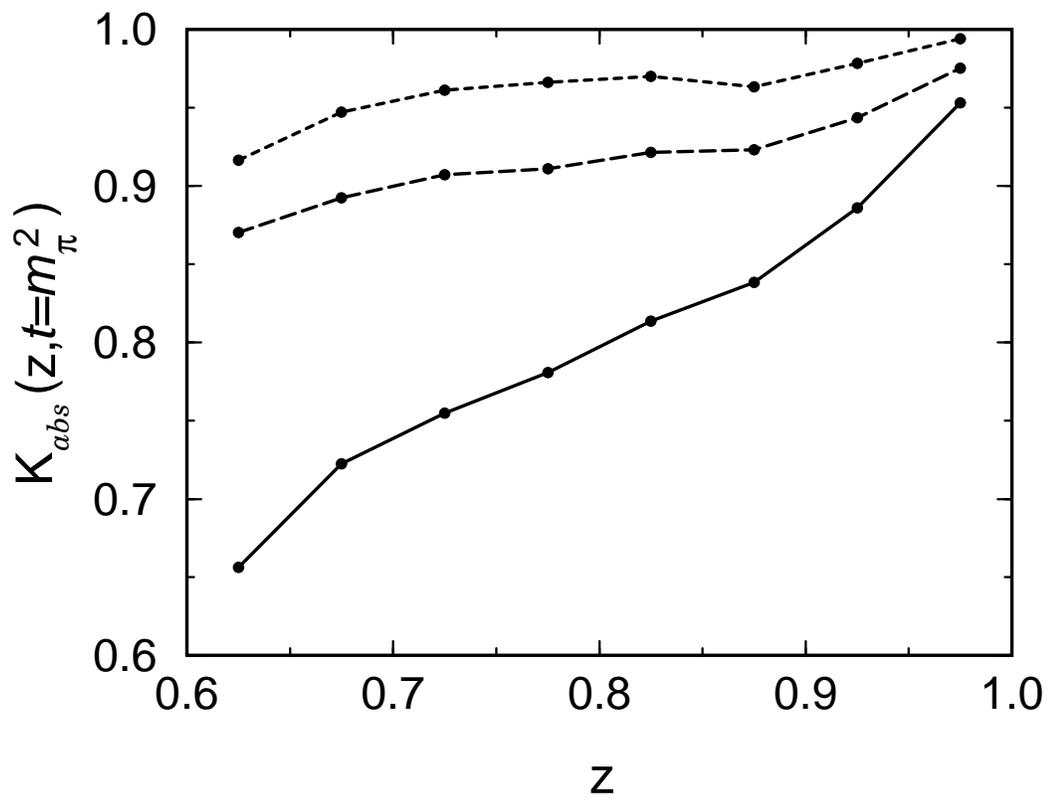
\end{document}